\documentclass[9pt,conference]{IEEEtran}
\IEEEoverridecommandlockouts
\usepackage{amsmath,amssymb,amsfonts}
\usepackage{comment}
\usepackage{cite}
\usepackage{graphicx}
\usepackage{textcomp}
\usepackage[breaklinks,colorlinks,linkcolor=blue,citecolor=blue,urlcolor=blue]{hyperref}
\usepackage{colortbl}
\usepackage{multirow}
\usepackage{footnote}
\usepackage{tablefootnote}
\usepackage{float}
\usepackage{threeparttable}
\usepackage{pifont}
\usepackage{textcomp}
\usepackage{tabularx}
\usepackage[table]{xcolor}
\usepackage{placeins}

\usepackage{algorithmic}
\usepackage[ruled,vlined,linesnumbered,commentsnumbered]{algorithm2e}
\usepackage{arydshln}

\def\BibTeX{{\rm B\kern-.05em{\sc i\kern-.025em b}\kern-.08em
    T\kern-.1667em\lower.7ex\hbox{E}\kern-.125emX}}

\begin{document}

\newcommand{\ml}[1]{{\color{red}\bf [Meng: #1]}}
\newcommand{\yx}[1]{{\color{blue}\bf [Yixuan: #1]}}
\newcommand{\zjw}[1]{{\color{green}\bf [zjw: #1]}}
\newcommand{\xt}[1]{{\color{black} #1}}
\newcommand{\xietong}[1]{{\color{purple}\bf [xt: #1]}}
\newcommand{\zw}[1]{{\color{magenta}[Zishen: #1]}}

\newcommand{\red}[1]{{\color{red}\bf (#1)}}
\newcommand{\magenta}[1]{{\color{magenta} #1}}
\newcommand{\method}{ReaLM}

\title{\method: \underline Reliable and \underline Efficient Large \underline Language \underline Model Inference with Statistical \underline Algorithm-Based Fault Tolerance \vspace{-15pt}}

\author{
Tong Xie$^{21}$,
Jiawang Zhao$^{1}$,
Zishen Wan$^{5}$,
Zuodong Zhang$^{24}$,
Yuan Wang$^{23}$,
Runsheng Wang$^{234}$,
Ru Huang$^{234}$,
and Meng Li$^{123*}$
\\
\textit{$^1$Institute for Artificial Intelligence \& $^2$School of Integrated Circuits, Peking University, Beijing, China} \\
\textit{$^3$Beijing Advanced Innovation Center for Integrated Circuits, Beijing, China} \\
\textit{$^4$Institute of Electronic Design Automation, Peking University, Wuxi, China} \\
\textit{$^5$Georgia Institute of Technology, Atlanta, GA, USA}

\vspace{-20pt}
\thanks{

$^*$Corresponding author: meng.li@pku.edu.cn

This work was supported in part by National Natural Science Foundation of China under Grant 62495102 and Grant 92464104, in part by Beijing Municipal Science and Technology Program under Grant Z241100004224015, and in part by 111 Project under Grant B18001.}
}

\vspace{-20pt}


\newcommand{\opt}{\texttt{OPT-1.3B}}
\newcommand{\llama}{\texttt{LLaMA-2-7B}}
\newcommand{\llm}{\texttt{LLaMA-3-8B}}

\maketitle
\begin{abstract}

The demand for efficient large language model (LLM) inference has propelled the development of dedicated accelerators. As accelerators are vulnerable to hardware faults due to aging, variation, etc, existing accelerator designs often reserve a large voltage margin or leverage algorithm-based fault tolerance (ABFT) techniques to ensure LLM inference correctness. However, previous methods often overlook the inherent fault tolerance of LLMs, leading to high computation and energy overhead. To enable reliable yet efficient LLM inference, in this paper, we propose a novel algorithm/circuit co-design framework, dubbed~\method. For the first time, we systematically characterize the fault tolerance of LLMs by performing a large-scale error injection study of representative LLMs and natural language understanding tasks. Then, we propose a statistical ABFT algorithm that fully leverages the error robustness to minimize error recovery as much as possible. We also customize the error detection circuits to enable a low-cost online collection of error statistics. Extensive experiments show that 
with \xt{only 1.42\% circuit area and 1.79\% power overhead, our ReaLM can reduce perplexity degradation from 18.54 to 0.29. Compared to existing methods},
ReaLM consistently reduces recovery costs across different operating voltages and improves energy efficiency by up to 35.83\% without compromising LLM performance. 
Our error injection code is available at \url{https://github.com/PKU-SEC-Lab/ReaLM_DAC25/}
\end{abstract}


\vspace{-10pt}
\section{Introduction}
Large language models (LLMs) have demonstrated remarkable performance across various tasks such as natural language processing \cite{brown2020language}, text generation \cite{zhao2023survey}, etc.
With trillions of operations, LLMs are widely deployed on customized accelerators such as TPU-like systolic arrays (SAs) \cite{jouppi2017datacenter, markidis2018nvidia} to meet their enormous computational demands.
However, due to aging, process variation, noise, etc \cite{dixit2021silent, moghaddasi2023dependable, jiao2017clim, huang2017variability}, reliability issues, e.g., timing violations, may emerge and make it challenging to ensure the computational correctness on these accelerators. Therefore, how to enable reliable yet efficient LLM acceleration becomes an important question and attracts increasing attention. 

\begin{figure}[!tb]
    \centering
    \includegraphics[width=1\linewidth]{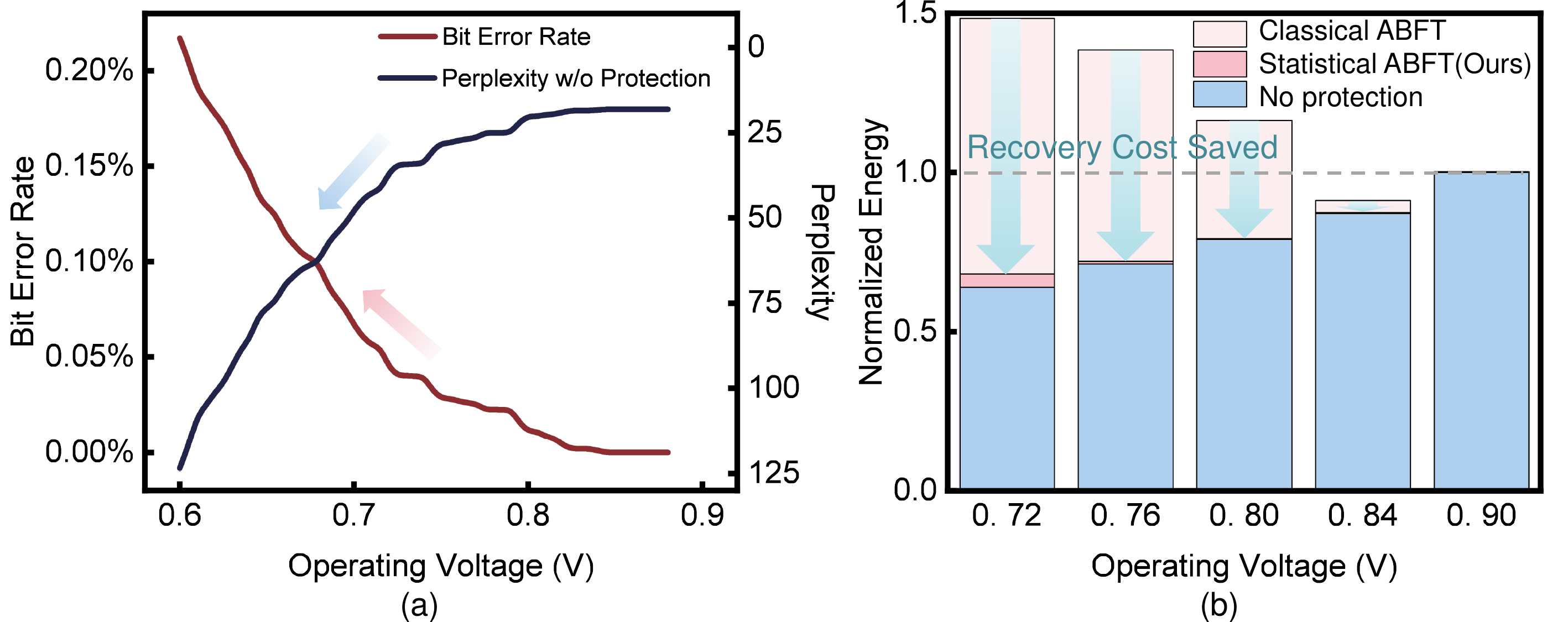}
    \vspace{-20pt}
    \caption{(a) Lower operating voltages increase BER, leading to significant perplexity degradation without protection. (b) Leveraging model resilience can reduce recovery costs. 
    BERs are synthesized on an SA with commercial 14nm PDK (nominal voltage: 0.9V), aligning with prior studies \cite{ernst2003razor,zhang2023read,wan2024mulberry}. Perplexity is evaluated using \opt~on WikiText-2 dataset. 
    }
    
    \label{fig:1}
    \vspace{-12pt}
\end{figure}

Although a large voltage margin can be reserved to prevent timing errors, it is usually not preferred due to the drastic increase in hardware costs. Meanwhile, directly reducing the voltage margin may raise the computation bit error rate (BER) and result in severe performance degradation, as shown in Fig.~\ref{fig:1}(a). To enable reliable yet efficient computation, various fault mitigation techniques from the circuit to the algorithm level have been proposed but often remain costly in the context of LLM inference. At the circuit level, techniques such as redundancy \cite{libano2018selective, khoshavi2020shieldenn}, and Razor flip-flops (FFs) \cite{ernst2003razor,whatmough2018dnn,zhang2018thundervolt,gundi2020effort} are developed for traditional logic circuits but usually lack scalability for deep learning accelerators. At the algorithm level, fault-aware fine-tuning has been suggested for conventional convolutional neural networks (CNNs) \cite{schorn2019efficient,kim2018matic,he2019noise}. However, the associated costs for LLM fine-tuning are prohibitive. Algorithm-based fault tolerance (ABFT) \cite{huang1984algorithm}, an algorithm and circuit co-design technique designed for general matrix-matrix multiplication (GEMM), emerges as a cost-effective solution for error detection. ABFT detects errors in the computation results and triggers error recovery mechanisms, e.g., overvolting in dynamic voltage frequency scaling (DVFS) \cite{cherupalli2016exploiting,constantin2015exploiting,jia201919}, recomputation, leading to much better efficiency.

However, as shown in Fig.~\ref{fig:1}(b), we observe that ABFT still faces efficiency challenges due to the high recovery costs. This is because classical ABFT triggers recovery for each detected computation error, which may not be necessary. As in Fig.~\ref{fig:1}(a), LLM performance does not degrade immediately at low BER, indicating inherited error resilience. By leveraging the error resilience, unnecessary recovery can be minimized to boost the system's energy efficiency. Nevertheless, the error resilience of LLMs is still not yet well understood. While past works have studied CNNs extensively \cite{jiao2017assessment,li2017understanding,reagen2018ares,mahmoud2021optimizing}, the complex architecture, diversified computation patterns, and unique quantization strategies of LLMs can make it significantly different.

Therefore, to fully exploit this inherent error resilience, we propose \method, an algorithm-circuit co-design framework that features reliable and efficient LLM inference. To the best of our knowledge, this is the first work that systematically and comprehensively characterizes the error resilience of LLMs. Our investigation reveals significant variations in error resilience across LLM network components due to the normalization operations. We also identify a trade-off between error magnitude and frequency in their impact on model performance. Building on these characterization insights, \method~further proposes a statistical ABFT approach to enable reliable and efficient LLM inference. \method~leverages the inherent resilience variations in LLMs to adaptively protect LLM network components in a cost-effective manner. 
We summarize our contributions as follows:
\begin{itemize}
    \item We perform a large-scale error injection study on representative LLMs and systematically characterize how various hardware faults affect LLM performance. Our characterization 
   reveals significant inherent resilience variations across LLM network components and identifies a trade-off between error magnitude and frequency.
    \item Based on the characterization insights, we propose a statistical ABFT strategy featuring an adaptive error correction \emph{algorithm} and a customized low-cost error detection \emph{circuit}. The proposed strategy integrates seamlessly with SA, minimizing recovery costs while maintaining performance. It further enables energy savings by leveraging LLM's inherent resilience characteristics.
    \item  Experimental results demonstrate that the proposed \method~significantly enhances error tolerance. ReaLM \xt{reduces perplexity degradation from 18.54 to 0.29 with only 1.42\% circuit area and 1.79\% power overhead}, preserving LLM performance while significantly reducing recovery costs and delivering up to 35.83\% energy savings compared to state-of-the-art methods.
    

\end{itemize}

\section{Background}

\subsection{Quantized Inference of Generative LLMs 
} 


\begin{figure}[!tb]
    \centering
    \includegraphics[width=0.9\linewidth]{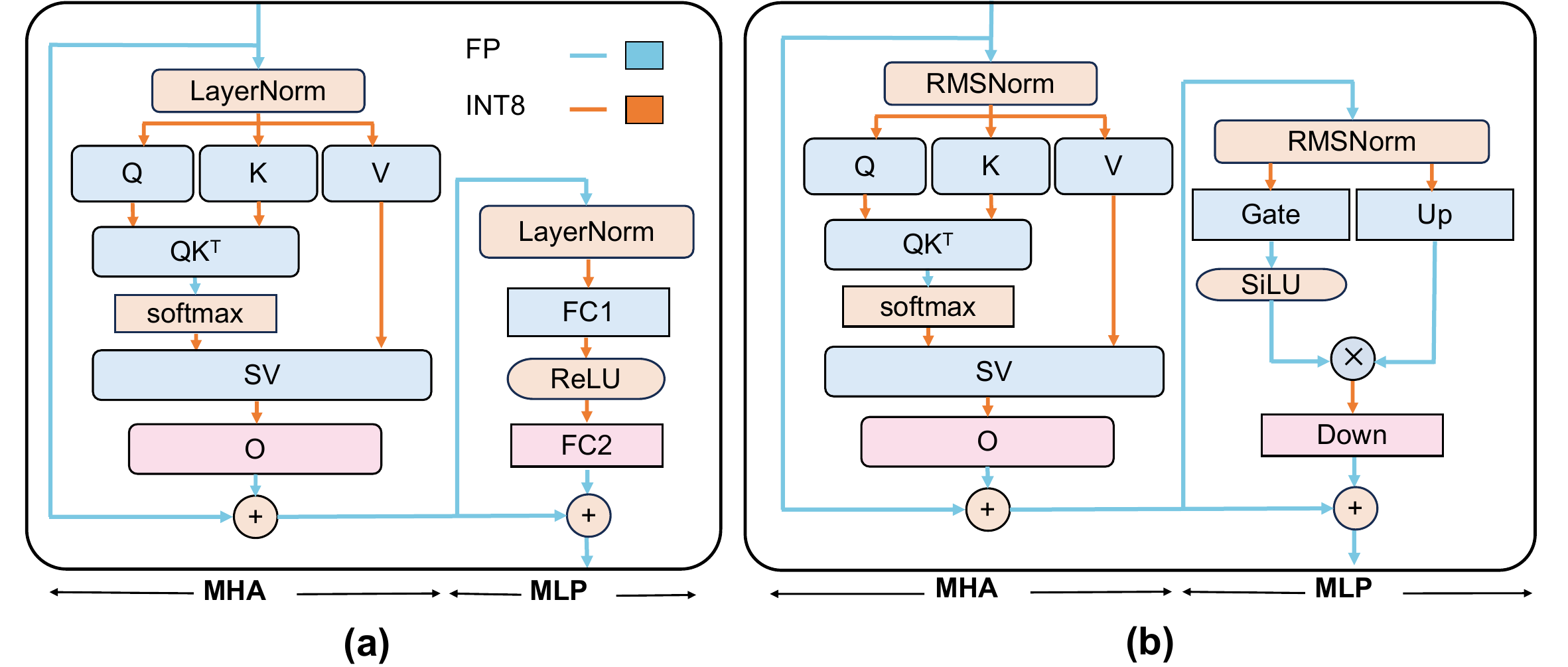}
    \vspace{-8pt}
    \caption{Transformer blocks of (a) \texttt{OPT} and (b) \texttt{LLaMA}.  
    }
    \label{fig:LLM}
    \vspace{-15pt}
\end{figure}

\label{sec:generative_inference}

A typical LLM architecture primarily consists of stacked Transformer blocks, each containing several GEMM operations and various nonlinear functions.
Fig. \ref{fig:LLM}(a) and (b) illustrate the architectures of two popular Transformer variants, namely 
the \texttt{OPT} block \cite{zhang2022opt} and the \texttt{LLaMA} block \cite{touvron2023llama}, respectively.
Both variants consist of various \textbf{network components}, as labeled in Fig. \ref{fig:LLM}, which will be referred to as \texttt{Q}, \texttt{Down}, and so on throughout this paper. 
Existing works~\cite{xiao2023smoothquant,lin2023awq,yuan2023rptq} have enabled low bit-width integer quantization of these GEMM operations, while nonlinear functions are retained in floating-point (FP) for accuracy.
The generative inference process of an LLM is divided into two stages: the prefill stage takes a prompt sequence as input, and the decode stage produce a single output token each time.
\subsection{Fault-Tolerant Deep Learning}
\label{fault_tolerant_DL}


\begin{table*}[t]
\centering
\footnotesize
\caption{Comparison of representative fault mitigation techniques.}
\label{tab:previous_work}
\renewcommand*{\arraystretch}{1.05}
\resizebox{\linewidth}{!}{%
\begin{tabular}{c|ccccccc} 
\hline \hline
\multirow{2}{*}{Method} & \multirow{2}{*}{Level} & Detection & {Hardware} & {Recovery} & {Recovery} & \multirow{2}{*}{Scalability} & {Compatibility w/} \\ 
& & Capability& Efficiency & Efficiency&Capability & & Accelerators\\ 
\hline \hline
Redundancy \cite{libano2018selective, khoshavi2020shieldenn} & circuit & high & low & low & high & medium & medium \\
\hline
Razor FFs \cite{ernst2003razor,whatmough2018dnn,zhang2018thundervolt,gundi2020effort} & circuit & high & low & medium & low & low & low \\
\hline
Fault-aware Fine-tuning \cite{schorn2019efficient,kim2018matic,he2019noise} & algorithm & - & - & prohibited & - & low & - \\
\hline
Classical ABFT \cite{huang1984algorithm, libano2023efficient, marty2020safe} & circuit-algorithm & high & medium & low & high & high & high \\
\hline
\rowcolor{gray!50}  Ours & circuit-algorithm & high & high & high & high & high & high \\
\hline \hline
\end{tabular}
\vspace{-30pt}
}
\end{table*}



\textit{Error sources.} 
Hardware accelerators are susceptible to various permanent and transient faults. Permanent faults often result from fabrication defects, while transient faults primarily arise from timing errors~\cite{salami2018resilience,jiao2017clim}. As CMOS technology scales down to the nanoscale, effects such as aging, variation, and noise become increasingly pronounced ~\cite{dixit2021silent, huang2017variability, moghaddasi2023dependable,jiao2017clim}. while voltage underscaling further reduces the design margin intended to mitigate these issues. Together, these factors lead to unintended timing delays, ultimately causing timing errors, as shown in Fig. \ref{fig:1}(a).




\textit{Model resilience.} 
To effectively characterize the inherent resilience of neural networks, many works \cite{sangchoolie2017one,he2020fidelity,hsiao2023silent,papadimitriou2021demystifying, zhang2023read} have employed a single-bit flip error model.
\cite{li2017understanding} analyzed fault propagation in networks, and
\cite{reagen2018ares} assessed how quantization improves fault tolerance.
\cite{mahmoud2021optimizing} identified vulnerability differences of certain feature maps
and suggested selective protection. 
\cite{wan2021analyzing, wan2024mulberry} further extended resilience studies to reinforcement learning applications. 
While \cite{agarwal2023resilience} explored error resilience in LLMs, they did not quantify the relationship between errors and model performance.
Considering the intricate model architecture, diversified network components, and unique quantization strategies, the error resilience of LLMs can be significantly different from traditional CNN models.


\textit{Error mitigation techniques.} 
Error correction code (ECC) \cite{hamming1950error} can effectively address hardware faults in memory or on-chip buffers. Mitigating transient faults in computational datapaths is more challenging\cite{hsiao2023silent,wan2021analyzing,hochschild2021cores,papadimitriou2021demystifying}, 
where various levels of protection can be leveraged, 
as summarized in Tab. \ref{tab:previous_work}.
However, these methods incur significant hardware or runtime overhead, particularly for LLMs.
For example, redundancy methods \cite{libano2018selective, khoshavi2020shieldenn} 
such as double-modular redundancy (DMR), require multiple copies of computations to be performed concurrently.
Circuit-level techniques~\cite{ernst2003razor,whatmough2018dnn,zhang2018thundervolt,gundi2020effort} integrate shadow FFs to detect timing errors but lack scalability in modern accelerators.
Algorithmic-level approaches~\cite{schorn2019efficient,kim2018matic,he2019noise} incorporate faults in fine-tuning to enhance error resilience. However, retraining LLMs is prohibitively computation-intensive.

\begin{figure}[!tb]
    \centering
    \includegraphics[width=0.9\linewidth]{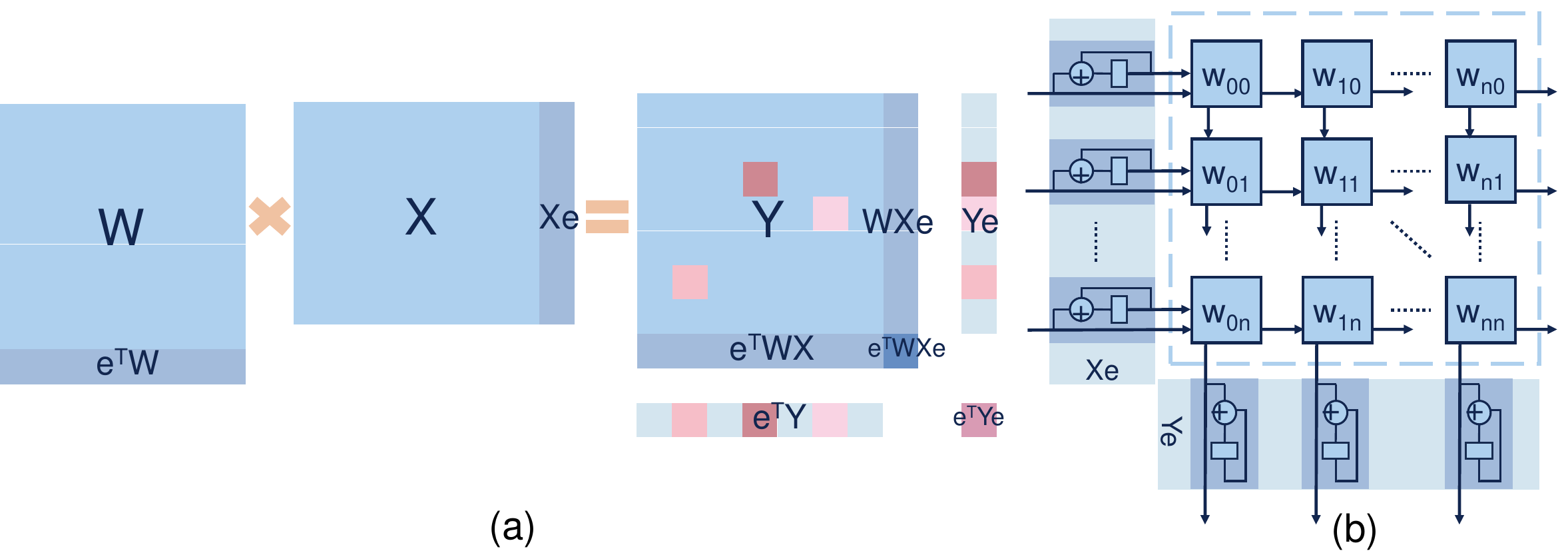}
    \vspace{-10pt}
    \caption{(a) Principle of ABFT. The checksums are compared to detect errors and capture error statistics. (b) Implementation of ABFT on SA \cite{bal2023novel}.}
    \label{fig:ABFT}
    \vspace{-16pt}
\end{figure}

\vspace{-5pt}
\subsection{Algorithm-Based Fault Tolerance (ABFT)}
\label{sec:ABFT}

ABFT leverages checksums to enhance reliability in GEMM \cite{huang1984algorithm}, as depicted in Fig. \ref{fig:ABFT}(a). In classical ABFT, matrices $W$ and $X$ are augmented with a checksum row $e^{\mathrm T} W$ and column $Xe$ for error detection, where $e$ is a unit column vector. 
Recent studies \cite{libano2023efficient, marty2020safe, xue2023approxabft, bal2023novel, safarpour2021algorithm} also propose various lightweight detection algorithms which only compute 
one-side checksums, i.e., $e^{\mathrm T} WX$ or $WXe$, or 
matrix sum deviation (MSD), i.e., $e^{\mathrm T} W Xe$, to lower ABFT hardware cost, as shown in Fig.~\ref{fig:ABFT}(a). We also show an example of integrating ABFT with the SA following \cite{bal2023novel} in Fig.~\ref{fig:ABFT}(b).
While ABFT is suitable for LLMs that depend extensively on GEMM operations, it relies on precise checksum-based comparisons and overlooks the inherent resilience of neural networks.
Recent work ApproxABFT \cite{xue2023approxabft} uses MSD to assess the error significance and allow small computational errors. However, this method ignores the impact of error frequency on the model accuracy and can lead to unnecessary recomputations.


Therefore, to address these challenges, we introduce \method, a circuit-algorithm co-design framework designed to enhance the reliability and efficiency of LLM inference.
We first inject computational errors into LLMs to study their resilience characteristics. Based on this characterization, we determine an adaptive error detection strategy to avoid unnecessary error recovery. Guided by this strategy, we customize a low-cost circuit to capture the computational error statistics.

\vspace{-12pt}
\section{Error Injection Framework}


To study the resilience of LLMs, we performed statistical error injection experiments, the most widely used approach for assessing the effects of hardware faults on various models.

\subsection{Error Model}
\label{subsec:error_model}
Given that (i) permanent errors are relatively straightforward to detect~\cite{isermann1984process} and do not require mitigation during runtime, and 
(ii) hardware faults in memory can be effectively addressed using ECC \cite{hsiao2023silent,wan2021analyzing,hochschild2021cores,papadimitriou2021demystifying,mahmoud2021optimizing}, 
in this work, we focus on transient computational errors occurring during inference and assume that data is correctly fetched from memory elements.
To model hardware faults, we adopt the commonly used \xt{random} bit-flip model~\cite{sangchoolie2017one,he2020fidelity,hsiao2023silent,papadimitriou2021demystifying, zhang2023read} as an abstraction.
Specifically, timing errors are modeled by injecting errors to higher bits, since they often affect more significant bits \cite{jiao2017clim,zhang2023read,safarpour2021algorithm}.
The severity of transient faults can be controlled by varying bit error rates.

\vspace{-5pt}
\subsection{Error Injection Method}

We conduct our error injection study on two representative LLMs: \opt~\cite{zhang2022opt} and \llama~\cite{touvron2023llama}. 
To simulate computational errors, we developed a dynamic error injection simulation framework using PyTorch, wherein bit flips
are emulated as tensor operations 
during runtime. 
Following \cite{xiao2023smoothquant}, the inputs of GEMM are quantized to INT8, while the results are in INT32 format.
Errors are then injected into the INT32 results of GEMM to realistically emulate compute errors in accelerators, which aligns with previous works \cite{sangchoolie2017one,he2020fidelity,hsiao2023silent,papadimitriou2021demystifying}.

Furthermore, to discern the different impacts of individual large errors versus multiple smaller errors that yield the identical MSD, we inject identical errors with defined {error} magnitude ($mag$) and {error} frequency ($freq$), adhering to the relationship $MSD = freq \times mag$. 
It is important to note that while this method diverges from the error models discussed in Sec. \ref{subsec:error_model}, its effectiveness in assessing the resilience of LLMs is preserved. 

\vspace{-5pt}
\subsection{Performance Benchmarks}

To assess the effects of error injection, we also construct a comprehensive benchmark framework. 
Consistent with previous studies \cite{xiao2023smoothquant,lin2023awq,yuan2023rptq}, we evaluate model performance using the LAMBADA commonsense QA benchmark \cite{paperno2016lambada} and the WikiText-2 language modeling task \cite{merity2016pointer}. 
Performance on LAMBADA is measured by accuracy ($\uparrow$), while WikiText-2 is evaluated using perplexity ($\downarrow$) \cite{jelinek1977perplexity}.
{We further capture the autoregressive generation capabilities of LLMs using the X-Sum\cite{narayan2018don} text summarization task, measured by ROUGE-1\cite{lin2004rouge} scores ($\uparrow$), and the GSM8K\cite{cobbe2021training} arithmetic reasoning task, evaluated by accuracy ($\uparrow$).} Our framework can be extensible to other benchmarks, like Hellaswag \cite{zellers2019hellaswag} (measured by accuracy ($\uparrow$)), which we examine in the evaluation section.

\section{LLM Resilience Characterization}
\label{sec:resilience}
In this section, we present our LLM resilience characterization in detail, setting out to answer the following research questions:
\begin{itemize}
    \item \textbf{Q1.1}: How does resilience vary across different layers of LLMs? (Sec. \ref{sec: layer})
    \item \textbf{Q1.2}: What is the bit-wise resilience of LLMs? (Sec. \ref{sec:bit})
    \item \textbf{Q1.3}: How does resilience vary among different components within LLMs during the prefill stage? (Sec. \ref{sec:component}) 
    \item \textbf{Q1.4}: What is the correlation between error magnitude and frequency in impacting LLM performance? (Sec. \ref{sec: tradeoff})
    \item \textbf{Q2.1}: How does the resilience of LLMs compare between the prefill stage and the decode stage? (Sec. \ref{sec:decode})
    \item \textbf{Q2.2}: How does resilience vary among different computational components within LLMs during the decode stage? (Sec. \ref{sec: decode_component}) 
\end{itemize}

To control for irrelevant variables, we implement the following protocols across research questions. 
(1) For Q1.4, we exclusively use \opt~model for fast assessment, whereas for Q2.1 and Q2.2, we employ only \llama~model to ensure reasonable performance. For all other questions, both models are utilized.
(2) Error injections are applied to all layers for all questions, except for Q1.1. 
(3) In Q1.1, Q1.3, Q2.1 and Q2.2, we specifically target the 30th bit for flipping. 
(4) For Q1.1 and Q2.1, errors are injected across all network components within a single Transformer block simultaneously.

\subsection{Resilience Study on Prefill Stage}
In this subsection, we will directly give the conclusion of {Q1.1} and {Q1.2} since they are relatively straightforward, and we focus on discussing {Q1.3} and {Q1.4}.

\subsubsection{{Layer-wise resilience}}
\label{sec: layer}

\begin{figure*}[!tb]
    \centering
    \includegraphics[width=1\linewidth]{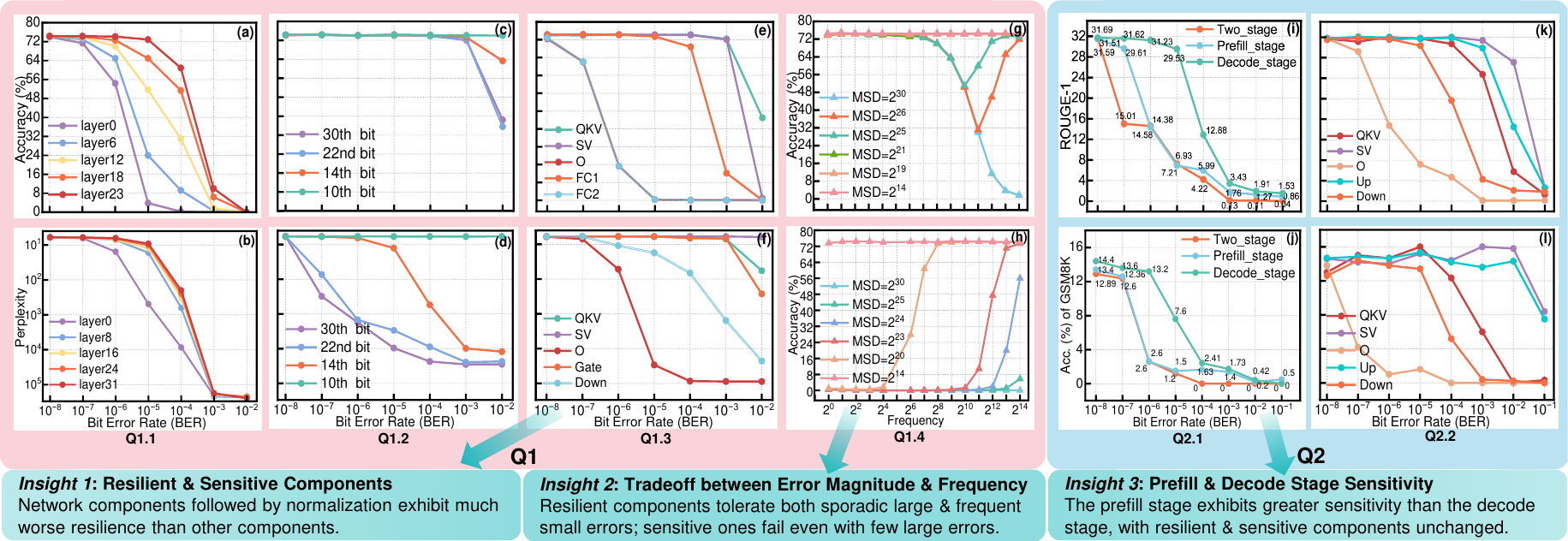}
    \vspace{-15pt}
    \caption{\textbf{Q1.1}: (a)(b) Layer-wise resilience of different LLMs on different tasks. 
    \textbf{Q1.2}: Bit-wise error resilience. (c) Error injection on \texttt{K}. (d) Error injection on \texttt{O}.  
    \textbf{Q1.3}: (e)(f) Sensitivity to errors in different LLM components. 
    \textbf{Q1.4}: Relationship between error frequency and magnitude. (g)  Resilient components like \texttt{K}. (h) Sensitive components like \texttt{O}. Given MSD, the error magnitude decreases as the error frequency increases. 
    \textbf{Q2.1}: (i)(j) Comparison between the prefill stage and decode stage.
    \textbf{Q2.2}: (k)(l) Impact of error injection across network components: \texttt{O}. and \texttt{Down} remain highly sensitive. 
    (a)(c)(e)(g)(h) are evaluated with \opt~on LAMDABA; (b)(d)(f) with \llama~on WikiText-2; (i)(k) with \llama~on X-Sum; (j)(l) with \llama~on GSM8K.     
    }
    \label{fig:resilience characterization}
    \vspace{-5pt}
\end{figure*}

As shown in Fig. \ref{fig:resilience characterization}(a)(b),
\textbf{while the layers exhibit comparable resilience behaviors, the earlier layers are more vulnerable to errors.} 



\subsubsection{Bit-wise resilience}
\label{sec:bit}

{By comparing error impacts in network components with FP outputs (\texttt{O}, Fig. \ref{fig:resilience characterization}(d)) and re-quantized INT8 outputs (\texttt{K}, Fig. \ref{fig:resilience characterization}(c)), we conclude that}
\textbf{while errors at lower bits have a negligible impact on model performance, errors at higher bits can reach a saturation point due to re-quantization.}



\subsubsection{Resilience of different network components}
\label{sec:component}



As depicted in Fig. \ref{fig:resilience characterization} (e)(f), our results reveal substantial differences in resilience behavior across the network components. Specifically, the \texttt{O} component in both models, \texttt{FC2} in \opt, and \texttt{Down} in \llama~exhibit significantly higher sensitivity compared to other components. For ease of discussion, these components are referred to as \textbf{\textit{sensitive components}}, while the others are categorized as \textbf{\textit{resilient components}}.

\begin{figure}[!tb]
    \vspace{-10pt}
    \centering
    \includegraphics[width=1\linewidth]{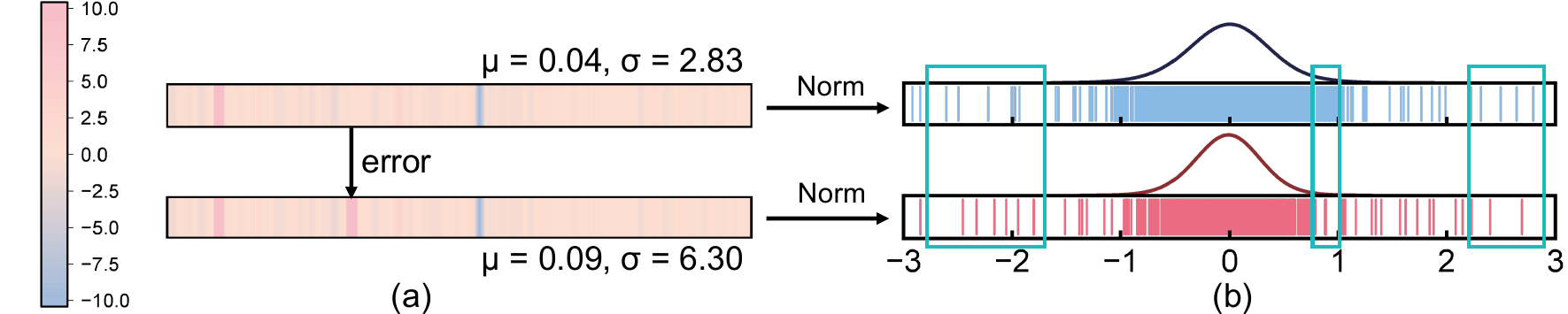}
    \vspace{-15pt}
    \caption{{(a) Data distribution of the pre-norm layer in LLMs, where outliers dominate $\mu$ and $\sigma$. Injecting larger errors can cause significant skew. (b) Data distribution after normalization is largely affected by the injected error.}}
    \label{fig:normalization}
    \vspace{-10pt}
\end{figure}

\textit{Discussion:} As demonstrated in Fig. \ref{fig:LLM}, the sensitive components in both \opt~and \llama~are followed by normalization layers—LayerNorm and RMSNorm, respectively,
unlike their resilient counterparts.
We demonstrate in Fig.~\ref{fig:normalization} that upon introducing a single error into the pre-normalization hidden states, a dramatic alteration in the subsequent elements is observed. This substantial variation is primarily due to the calculations of mean ($\mu$) and standard deviation ($\sigma$) during normalization.
In LLMs, hidden states inherently consist of a few outliers \cite{xiao2023smoothquant, lin2023awq, yuan2023rptq}, while the majority of elements remain close to zero. These outliers predominantly influence the computation of $\mu$ and $\sigma$, shaping the overall statistics of the hidden states.
Therefore, a larger error can manifest as an outlier and drastically skew these parameters, markedly altering the normalization outcome. In contrast, injecting a larger error into the resilient components affects only a localized set of elements. 
Thus, we conclude that \textbf{the network components followed by normalization exhibit much worse resilience than other components due to normalization.}
Specifically, for \texttt{QK}$^\texttt{T}$ components followed by softmax, changes remain confined because softmax outputs between 0 and 1.

\subsubsection{Tradeoff between error magnitude and frequency}
\label{sec: tradeoff}



While MSD can serve as a metric to depict the error severity, it cannot differentiate between the impacts of a single large error and multiple smaller errors, despite them potentially yielding identical MSD values and having significantly different effects on model performance.
Consequently, this section introduces additional metrics—error magnitude ($mag$) and error frequency ($freq$)—to more accurately represent error statistics. 
We define $mag$ as $MSD/freq$ in Fig. \ref{fig:resilience characterization}(g)(h), indicating that an increase in $freq$ results in a decrease in $mag$. Our analysis reveals that resilient components exhibit non-monotonic resilience behavior: both sporadic large errors and frequent smaller errors minimally impact model performance. Conversely, a moderate frequency of medium-sized errors can severely impair performance. However, for sensitive components, acceptable performance is maintained only when numerous smaller errors are introduced.


\textit{Discussion.} 
The non-monotonic resilience behavior observed in resilient components illustrates a trade-off between error magnitude and frequency. 
\textbf{Resilient components are tolerant to both sporadic large errors and frequent smaller ones, whereas sensitive components falter even with few large errors}. This finding aligns with earlier observations from Sec. \ref{sec:component}. Fig. \ref{fig:ABFT_strategy} further delineates a critical region where significant performance degradation occurs.

\vspace{-6pt}
\subsection{Resilience Study on Decode Stage}
\vspace{-4pt}
LLMs are primarily different from CNNs due to their autoregressive decode stage. In this section, we explore the resilience characterization of the decode stage and differentiate it from the prefill stage.

\subsubsection{Difference in prefill stage and decode stage}
\label{sec:decode}


To differentiate resilience between the prefill and decode stages, we inject errors exclusively into the prefill stage, the decode stage, and both stages, respectively. As illustrated in Fig. \ref{fig:resilience characterization}(i)(j), our analysis indicates that performance degradation is consistently more severe during the prefill stage than during the decode stage across both tasks.

\textit{Discussion.} \textbf{The prefill stage exhibits greater sensitivity to errors than the decode stage}. The reason lies in the KV-cache mechanism.
During the decode stage, the newly generated query vector interacts with the KV-cache, which is primarily composed during the prefill stage. Consequently, errors during the prefill stage compromise most of the operations involved in generating the next token. In contrast, errors during the decode stage have a lesser impact, as the largely unaffected KV cache drives most of computations for the next token generation.

\subsubsection{Resilience across compute units}
\label{sec: decode_component}


Similar to the experiments described in Sec. \ref{sec:component}, we inject errors into various LLM network components and present the outcomes in Fig. \ref{fig:resilience characterization}(k)(l). Our findings confirm that \textbf{the sensitive components, specifically \texttt{O} and \texttt{Down}, continue to be highly vulnerable}, echoing observations from Sec.~\ref{sec:component}. 

\section{Statistical Algorithm-Based Fault Tolerance}

Building on the insights from Sec. \ref{sec:resilience}, this section introduces our statistical ABFT framework. \xt{We integrate it with SAs, focusing on weight stationary (WS) and output stationary (OS) dataflows due to page limitations, though it is also compatible with input stationary (IS) dataflow.}


\subsection{Statistical-Based Error Detection Strategy}
\begin{figure}
    \centering
    \includegraphics[width=1\linewidth]{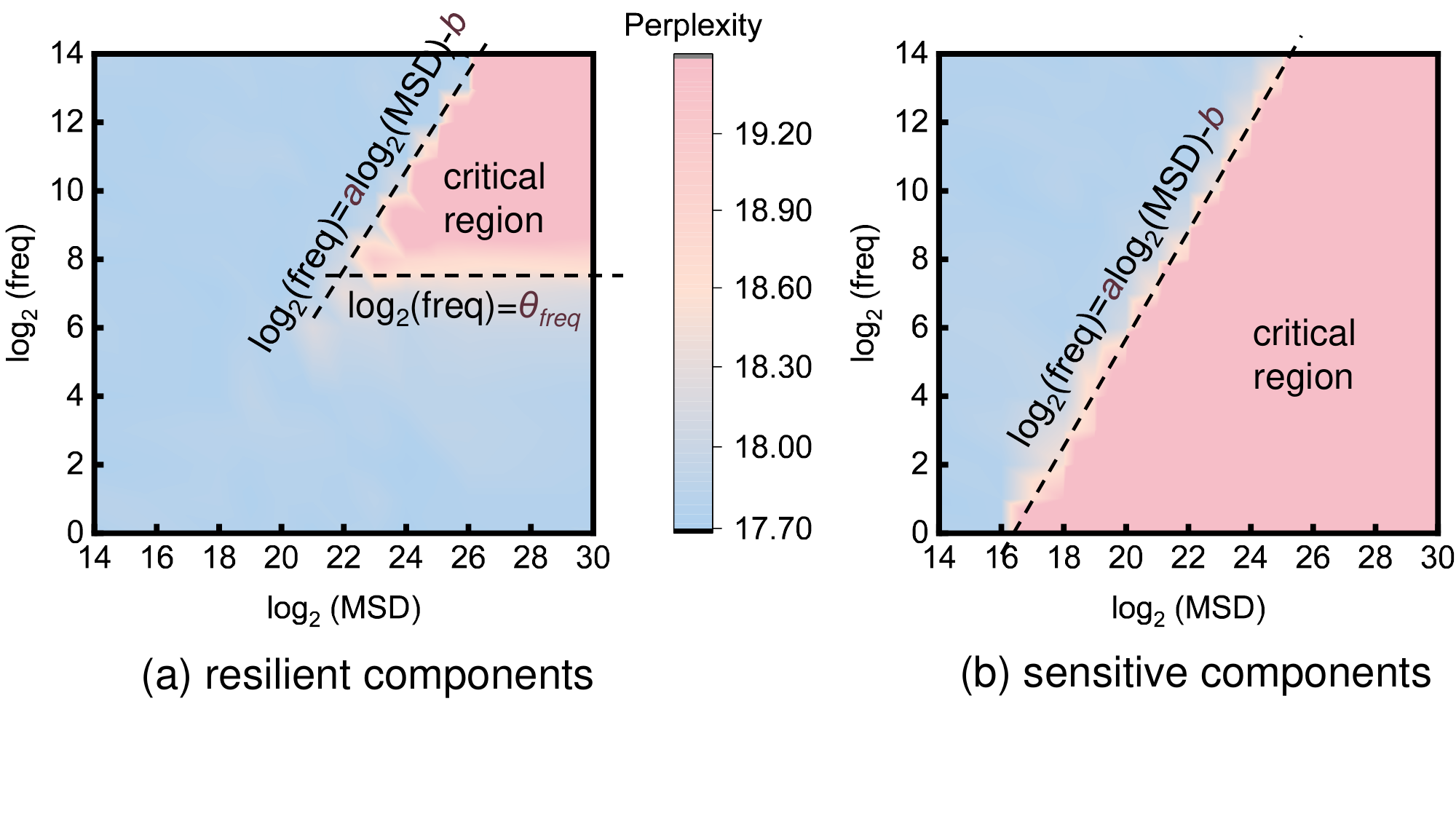}
    \vspace{-40pt}
    \caption{Our statistical ABFT strategy only corrects errors falling inside the critical region.}
    \vspace{-10pt}
    \label{fig:ABFT_strategy}
    
\end{figure}

\method~utilizes MSD, error frequency ($freq$), and error magnitude ($mag$) as key statistics to characterize the error distributions. 
As shown in Fig. \ref{fig:ABFT_strategy}, an \textit{acceptable} performance threshold can be empirically established to define the critical region where computational errors significantly impact LLM performance. \method~simply ignores errors outside the region. 


In terms of the critical region, we find that for resilient components, the boundary of the critical error region is defined by a horizontal line and an inclined line with a slope greater than 1. The horizontal line suggests that resilient components are tolerable to errors below a frequency threshold $\theta_{freq}$, i.e., the vertical coordinate of this line, regardless of their magnitude.
For sensitive components, the critical region is bounded by the inclined line and horizontal axis.
{Note that the slope $a>1$, the intercept $-b$ of the inclined line, and frequency threshold $\theta_{freq}$ are all fitted parameters.}

We derive several key conclusions to guide our strategy: 1) the horizontal bound indicates that errors exceeding a certain MSD or magnitude all lead to unacceptable performance degradation at $\theta_{freq}$. Therefore, errors larger than the threshold can be counted without distinction; 2) Errors positioned above the inclined line are deemed acceptable. This boundary delineates a clear threshold $\theta_{mag}$ beneath which errors, regardless of their frequency, do not significantly degrade the model performance; 3) Assuming that errors are equal, we can derive the upper bound of $\theta_{mag}=b-(a-1)\log_2{MSD}$. Errors smaller than $\theta_{mag}$ have negligible impact on model performance.

Based on these insights, we assess the influence of errors with error statistics in real-time. 
Our methodology ignores errors smaller than $\theta_{mag}$ due to their minimal impact on performance. 
Instead, we only count significant errors to calculate the effective error frequency $freq_{eff}=\mathrm{countif}(mag>\theta_{mag})$. 
We then compare $freq_{eff}$ to the threshold $\theta_{freq}$; if $freq_{eff}>\theta_{freq}$, the recovery is triggered. Otherwise, the GEMM operation is considered safe.

\vspace{-7pt}
\subsection{Architecture Design of Statistical ABFT}
\vspace{-3pt}

\begin{figure}[!tb]
    \centering
    \includegraphics[width=0.8\linewidth]{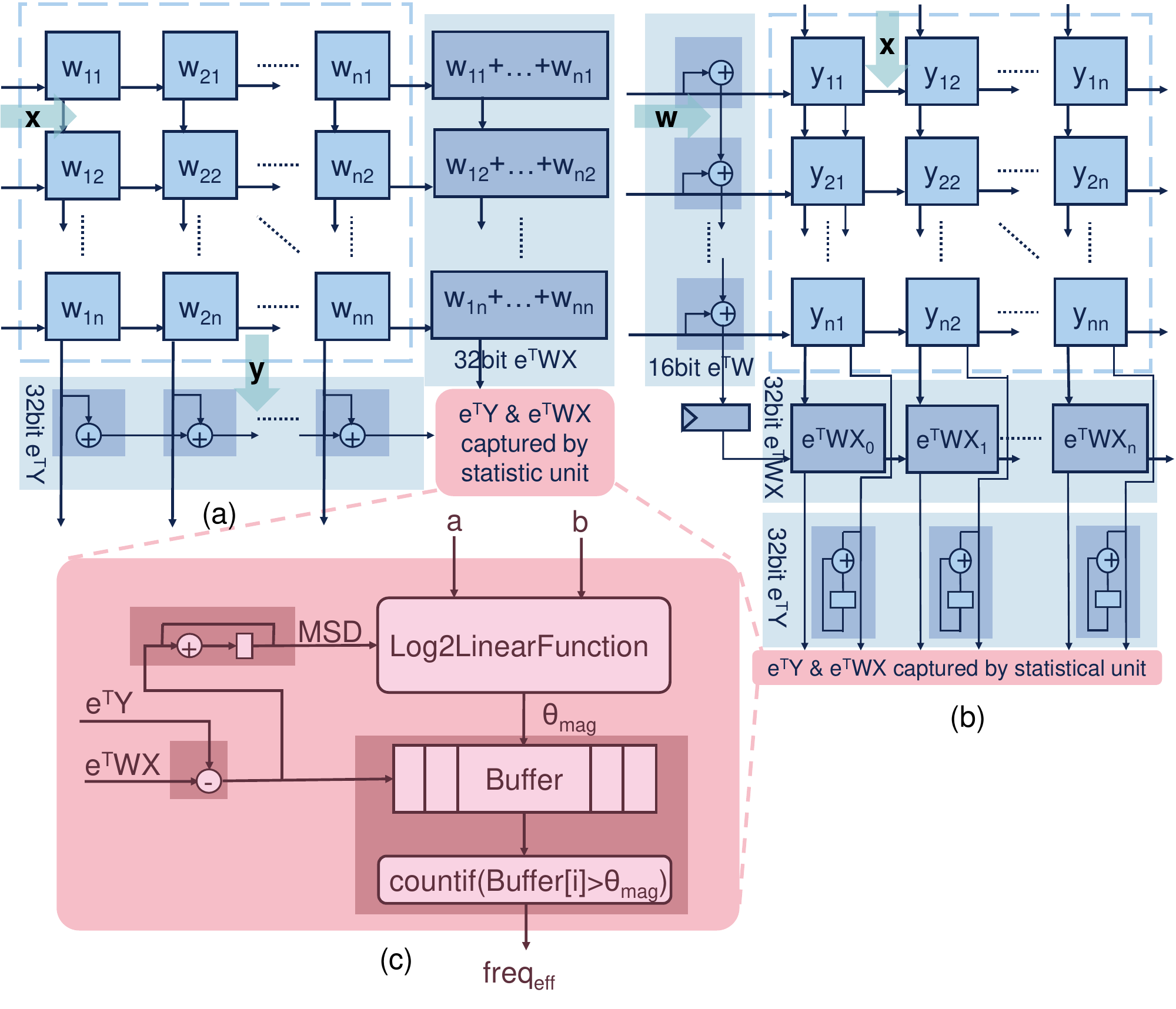}
    \vspace{-10pt}
    \caption{Architecture design of statistical ABFT on SA: (a) ABFT implementation for WS dataflow; (b) ABFT implementation for OS dataflow; and (c) customized statistical units.}
    \label{fig:architecure_design}
    \vspace{-10pt}
\end{figure}

We integrate our statistical ABFT into SA architecture and support both WS and OS dataflow (Fig.~\ref{fig:architecure_design}). 
We only need to add an extra ``statistical unit'', as shown in Fig.~\ref{fig:architecure_design}(c), to capture the error statistics and determine the need for recovery, which incurs a very small area overhead.
For ease of explanation, we consider the example of computing a GEMM of matrices $W, X \in \mathbb{R}^{n \times n}$ on an $n \times n$ SA, resulting in $Y \in \mathbb{R}^{n \times n}$.



In the WS configuration of the SA (Fig.~\ref{fig:architecure_design}(a)), weights are stored within the processing elements (PEs), and inputs are streamed horizontally from left to right. 
The array computes partial sums that move vertically downward. 
{A column of PEs is added on the right, storing $e^{\mathrm T} W$ for computing $e^{\mathrm T} WX$.}
Meanwhile, a row of adders at the bottom accumulates $e^{\mathrm T} Y$.
{This design minimizes the need for higher bit-width multipliers, as seen in \cite{bal2023novel}, with only minimal area cost.}
The latency overhead of computing $e^{\mathrm T} WX$ and $e^{\mathrm T} Y$ is also negligible.


Similarly, for the OS dataflow (Fig. \ref{fig:architecure_design}(b)), weights traverse horizontally from left to right, while inputs flow vertically from top to bottom. 
Outputs are generated directly at the intersection of weights and inputs within the PEs. 
We follow previous works \cite{safarpour2021algorithm} for our implementation.
An additional column of adders is integrated on the left side of the array, dedicated to computing $e^{\mathrm T} W$. 
This checksum is then routed to an extra row of PEs positioned at the bottom and propagated rightward to calculate the product $e^{\mathrm T} W X$. 
The outputs $y$ are vertically accumulated across the PEs to derive the output checksum $e^{\mathrm T} Y$. This configuration only requires the additional row of PEs to support higher-bit multiplications. 

The proposed statistical unit comprises a subtractor, an accumulator, a Log2LinearFunction unit, and a sequence of $n$ buffers equipped with a comparator-based ``countif'' unit.
The Log2Linear-Function unit is designed to calculate $\theta_{mag}=b-(a-1)\log_2{MSD}$. 
Throughout the computation process, checksums are continuously fed into the statistical unit.
They are first subtracted and the resultant differences are accumulated to compute MSD, which is then stored in the buffers. 
Once the accumulator completes the MSD computation, the data is forwarded to the Log2LinearFunction unit, where $\theta_{mag}$ is computed. Ultimately, the ``countif'' unit evaluates all buffer elements in parallel and counts the frequency of errors exceeding $\theta_{mag}$ to determine $freq_{eff}$.



In the decode stage, general matrix-vector multiplication (GEMV) dominates the computation. While batched GEMVs can also be treated as GEMMs, to which our method applies, GEMVs during non-batched inference are typically executed in specialized vector processing units, which fall outside the primary focus of SAs and ABFT techniques discussed in this paper.

\xt{
Therefore, based on the insights in Fig.~\ref{fig:resilience characterization}, the proposed statistical ABFT can enhance LLM resilience by adaptively leveraging captured error statistics. This, in turn, unlocks significant energy-saving opportunities, such as voltage underscaling, even in the presence of errors (Fig. \ref{fig:1}), enabling both reliable and efficient LLM inference.
 }

\section{Evaluation}
\subsection{Experiment Setup}


We first use the lightweight \opt~measuring perplexity on WikiText-2 for rapid evaluation.
To demonstrate generality, we also extend the evaluation to the recently released \llm~model and a different task, i.e., HellaSwag \cite{zellers2019hellaswag}.
Timing errors are simulated by randomly flipping bits in the accumulation results (represented as 32-bit integers) following \cite{jiao2017clim,zhang2023read}. 
The relation between the BER and the operating voltage follows Fig.~\ref{fig:1}.
To determine the parameters for the proposed statistical ABFT, i.e., $a, b, \theta_{freq}$, we inject errors into LLMs for performance evaluation. We empirically set the parameters to allow a 0.3 perplexity increase and a 0.5\% accuracy decrease as the acceptable performance degradation.


The RTL for both our proposed and existing ABFT designs is implemented and integrated into SAs with 256 × 256 PEs supporting both WS and OS dataflows. These designs are synthesized using Synopsys Design Compiler with {commercial 14nm PDK, where the nominal operating voltage is 0.9V. The clock cycle is set to 500 ps, with a critical path delay of 439 ps, and circuit area and power are obtained from synthesis results. We conduct timing analysis using Synopsys PrimeTime and HSPICE to capture BERs at reduced voltages, considering the toggle rate during real LLM inference, which is in line with previous chip measurement data \cite{ernst2003razor, wan2024mulberry, zhang2023read}. Error recovery is assumed to be performed through recomputation at nominal voltage to ensure correctness, and the total energy consumption is evaluated accordingly.}
\vspace{-5pt}
\subsection{Circuit Overhead Comparison}
\begin{figure}[!tb]
    \centering
    \includegraphics[width=0.9\linewidth]{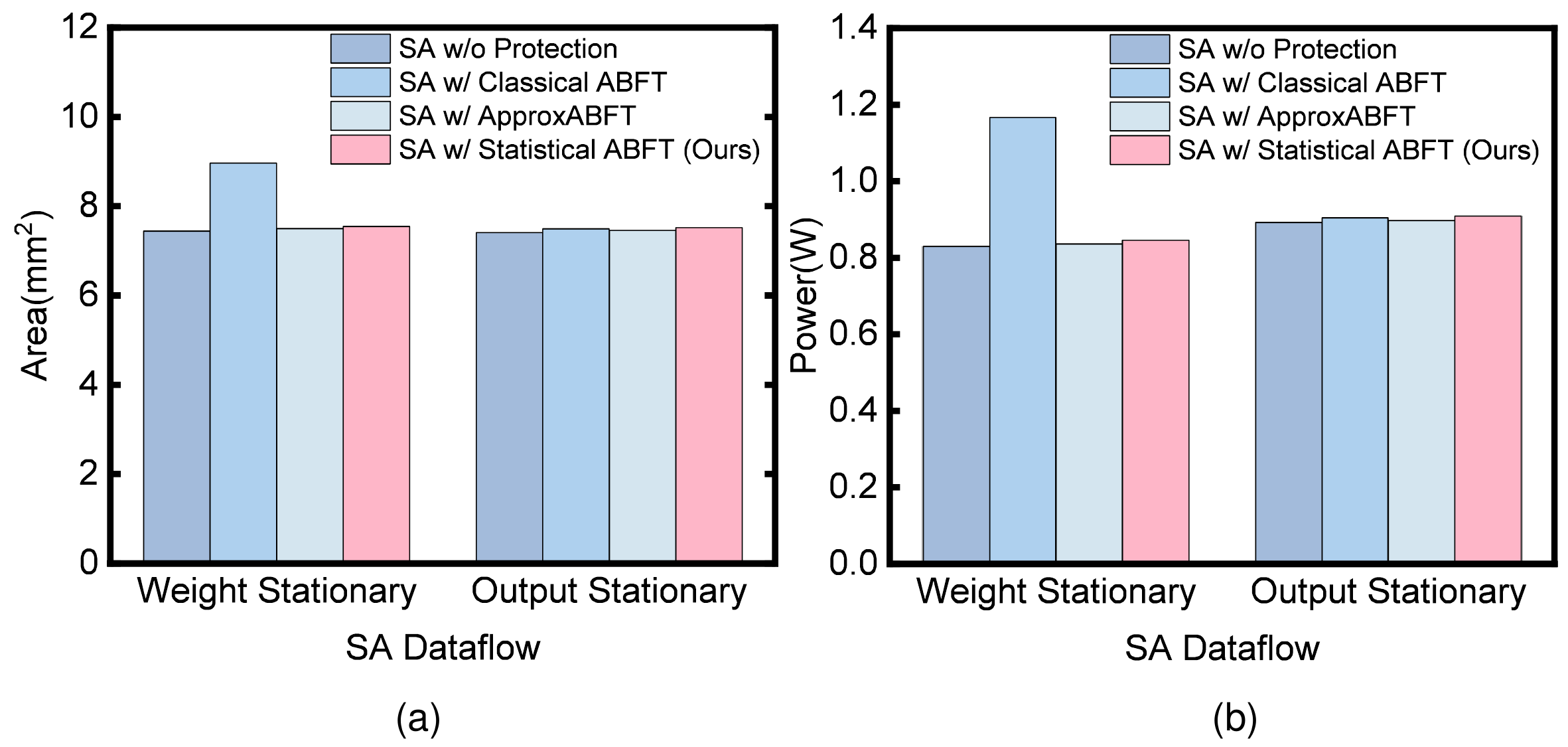}
    \vspace{-10pt}
    \caption{(a) Area for WS and OS dataflow; (b) Power for WS and OS dataflow.}
    \label{fig:circuit_overhead}
    \vspace{-5pt}
\end{figure}

\vspace{-3pt}
For both WS and OS dataflows, we analyze the circuit overhead in terms of area and power consumption across no {protection}, classical ABFT\cite{safarpour2021algorithm}, ApproxABFT\cite{xue2023approxabft}, and statistical ABFT, as depicted in Fig. \ref{fig:circuit_overhead}. 
\xt{Compared to unprotected SAs, our design introduces minimal overhead, with only 1.43\% area and 1.82\% power overhead for WS dataflow, and 1.42\% area and 1.79\% power overhead for OS dataflow.

}


\vspace{-4pt}
\subsection{LLM Performance and Energy Savings}
\vspace{-2pt}
\begin{figure}[!tb]
    \centering

    \includegraphics[width=1\linewidth]{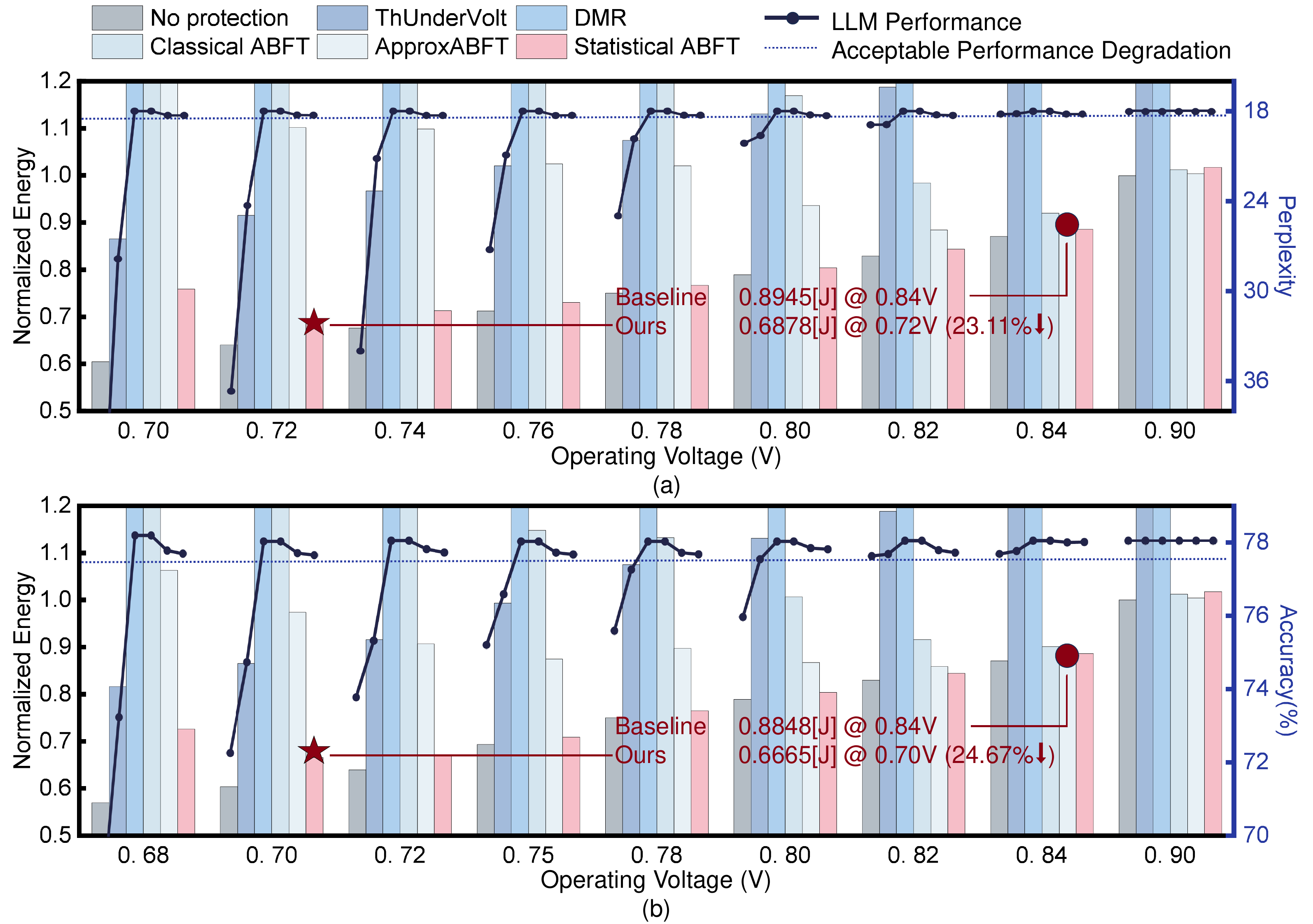}
    \vspace{-20pt}
    \caption{{LLM performance and total energy savings comparison. (a) \opt~on WikiText-2. (b) \llm~on HellaSwag.
    Our method preserves competitive performance with minimal protection overhead, 
    achieving maximal energy savings at 0.72V and 0.70V, respectively.
    }}   
    \label{fig: mainresults}
    \vspace{-10pt}
\end{figure}

{We apply our statistical ABFT to network component \texttt{K} in \opt~and \texttt{V} in \llm~as examples to demonstrate its effectiveness. Fig. \ref{fig: mainresults} compares the total energy overhead and LLM performance for different methods: no protection, ThunderVolt\cite{zhang2018thundervolt}, DMR, classical ABFT\cite{safarpour2021algorithm}, ApproxABFT\cite{xue2023approxabft}, and our statistical ABFT, across various operating voltages. 
Our statistical ABFT maintains comparable LLM performance while significantly reducing the recovery cost and therefore total energy.
For \texttt{K} in \opt, our approach achieves a sweet spot at 0.72V, \xt{reducing  perplexity degradation from 18.54 to 0.29 and}
saving 23.11\% energy compared to prior-art methods. Similarly, for \texttt{V} in \llm, it finds a sweet spot at 0.70V, \xt{lowering accuracy degradation from 7.64\% to 0.47\%} and saving 24.67\% energy.

}

\vspace{-7pt}
\subsection{Energy Savings across LLM Components}
\vspace{-3pt}

{We summarize 
the total energy savings
for all network components in Tab. \ref{tab:saving_across_components}. As shown, optimal operating voltages (i.e., sweet spots) and resulting benefits vary across components, reflecting the differences in resilience as discussed in Sec.~\ref{sec:resilience}. Our method achieves up to 35.83\% energy savings on \texttt{V} in \opt~and 34.46\% on \texttt{SV} in \llm. \xt{While resilient components demonstrate 
benefits, sensitive components (\texttt{O}, \texttt{FC2}, and \texttt{Down}) show more limited savings due to their vulnerability to errors, which reduces the potential to exploit inherent resilience.}

}

\begin{table}[!tb]
\centering
\caption{energy saving across network components}
\setlength{\tabcolsep}{4pt}
\begin{tabular}{ccc|ccc}
\hline \hline
\multicolumn{3}{c|}{{\opt}} & \multicolumn{3}{c}{{\llm}} \\ \hline
{Network} & {Optimal} & {Energy} & {Network} & {Optimal} & {Energy}\\
{Component} & {Voltage (V)} & {Saving} & {Component} & {Voltage (V)} & {Saving} \\ \hline \hline
\texttt{Q}   & 0.70  & 28.68\% & \texttt{Q}   & 0.71 & 24.26\% \\ 
\texttt{K}   & 0.72 & 23.11\% & \texttt{K}    & 0.72 & 24.17\% \\ 
\texttt{V}   & 0.65 & 35.83\% & \texttt{V}    & 0.70 & 24.67\% \\ 
\texttt{QK}$^{\texttt{T}}$  & 0.67 & 33.54\% & \texttt{QK}$^{\texttt{T}}$  & 0.73 & 19.46\% \\ 
\texttt{SV}  & 0.75 & 17.44\% & \texttt{SV}   & 0.66 & 34.46\% \\ 
\texttt{O}   & 0.82 & 3.38\%  & \texttt{O}    & 0.83 & 2.40\%  \\ \hline
\multirow{2}{*}{\texttt{FC1}} & \multirow{2}{*}{0.75} & \multirow{2}{*}{15.01\%} & \texttt{Gate} & 0.78 & 10.21\% \\ 
     &      &        & \texttt{Up}   & 0.77 & 16.56\% \\ 
\texttt{FC2} & 0.83 & 3.14\%  & \texttt{Down} & 0.83 & 3.12\%  \\ \hline \hline
\end{tabular}
\label{tab:saving_across_components}
\end{table}

\vspace{-4pt}
\subsection{Tradeoff between Performance and Energy Savings}
\vspace{-3pt}

\begin{figure}[!tb]
    \centering
    \includegraphics[width=0.9\linewidth]{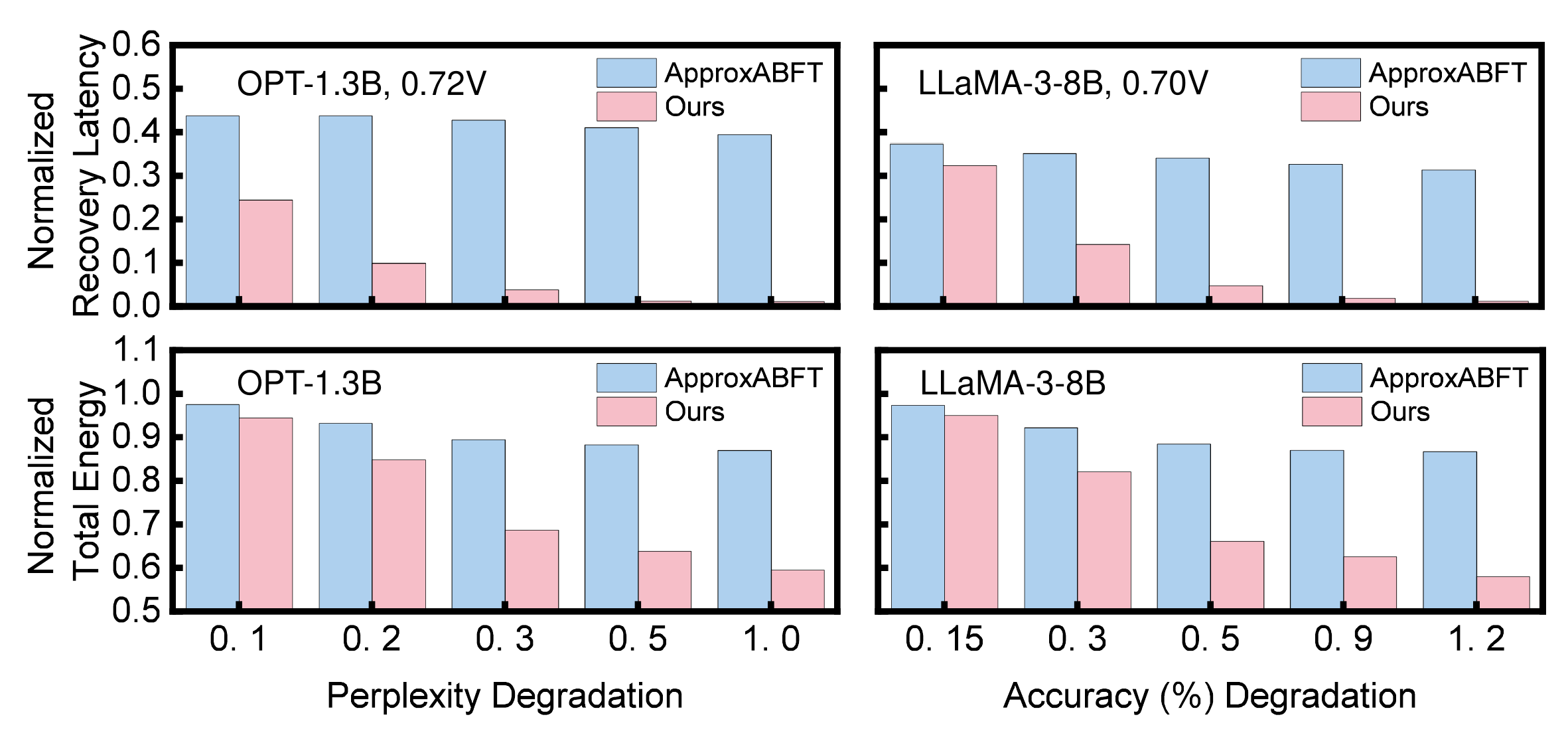}
    \vspace{-15pt}
    \caption{{Trade-off between acceptable performance degradation and its impact on recovery latency and total energy.}}   
    \label{fig: DSE}
    \vspace{-10pt}
\end{figure}

    {We explore the tradeoff between acceptable performance degradation and its impact on recovery latency and total energy over baseline~\cite{xue2023approxabft}  for \texttt{K} in \opt~and \texttt{V} in \llm, as shown in Fig.~\ref{fig: DSE}}. The recovery latency is evaluated at 0.72V and 0.70V, respectively, and the total energy is evaluated at the optimal voltages on each performance degradation constraint.
    This tradeoff demonstrates our design’s flexibility in balancing computational efficiency with reliability requirements.

\section{Conclusion}



In this paper, we introduce \method, an algorithm/circuit co-design framework aimed at enhancing the reliability and efficiency of LLM inference. We conduct a systematic error injection experiment to characterize the resilience of LLMs, highlighting particular vulnerabilities in the normalization operation that significantly affect inference reliability. Building on these insights, we develop a statistical ABFT approach featuring a customized, low-cost error detection circuit and an adaptive error correction strategy, effectively reducing error recovery costs. Our evaluations reveal that \xt{\method~can significantly enhance LLM performance by reducing perplexity degradation from 18.54 to 0.29 with only 1.42\% circuit area overhead and 1.79\% power overhead. By fully leveraging the inherent resilience of LLMs, \method~can achieve up to 35.83\% energy savings compared to existing methods while maintaining competitive model performance.

}


\newpage

\bibliographystyle{ieeetr}

\end{document}